\documentclass[10pt, journal]{IEEEtran}
\usepackage{url}
\usepackage[dvips]{color}
\usepackage{algpseudocode}
\usepackage{multirow}
\usepackage{mdframed}
\usepackage{lipsum}
\usepackage{setspace}
\usepackage{graphicx}
\usepackage{graphics}
\usepackage{latexsym}
\usepackage{showidx}
\usepackage{latexsym}
\usepackage{amssymb}
\usepackage{amsfonts}
\usepackage{amsmath}
\usepackage{amssymb}
\usepackage{amsfonts}
\usepackage{amsmath}
\newtheorem{Theorem}{Theorem}
\newtheorem{Proposition}{Proposition}
\newtheorem{Definition}{Definition}
\newtheorem{Problem}{Problem}

\newtheorem{lemma}{Lemma}
\newtheorem{rem}{Remark}

\newcommand{\blue}{\color{black}}

\usepackage[tight,footnotesize]{subfigure}
\usepackage{stfloats}
\begin{document}

\title{Integer-Forcing MIMO Linear Receivers Based on Lattice Reduction}
\author{\authorblockN{A.~Sakzad,~\IEEEmembership{Member,~IEEE}, J.~Harshan,~\IEEEmembership{Member,~IEEE}, and E.~Viterbo,~\IEEEmembership{Fellow,~IEEE}}\thanks{This work was performed at the Monash Software Defined Telecommunications (SDT) Lab and was supported by the Monash Professional Fellowship and the Australian Research Council under Discovery grants ARC DP 130100103.}\\ 
\authorblockA{Department of Electrical and Computer Systems Engineering,\\
Monash University, Melbourne, Victoria, Australia}\\
{\small {\tt $\{$amin.sakzad, harshan.jagadeesh, and emanuele.viterbo$\}$@monash.edu}}}
\and

\maketitle
\begin{abstract}
A new architecture called integer-forcing (IF) linear receiver has been recently proposed for multiple-input multiple-output (MIMO) fading channels, wherein an appropriate integer linear combination of the received symbols has to be computed as a part of the decoding process. In this paper, we propose a method based on Hermite-Korkine-Zolotareff (HKZ) and Minkowski lattice basis reduction algorithms to obtain the integer coefficients for the IF receiver. We show that the proposed method provides a lower bound on the ergodic rate, and achieves the full receive diversity. Suitability of complex Lenstra-Lenstra-Lovasz (LLL) lattice reduction algorithm (CLLL) to solve the problem is also investigated. Furthermore, we establish the connection between the proposed IF linear receivers and lattice {\blue reduction-aided} MIMO detectors (with equivalent complexity), and point out the advantages of the former class of receivers over the latter. For the $2 \times 2$ and $4\times 4$ MIMO channels, we compare the coded-block error rate and bit error rate of the proposed approach with that of other linear receivers. Simulation results show that the proposed approach outperforms the zero-forcing (ZF) receiver, minimum mean square error (MMSE) receiver, and the lattice {\blue reduction-aided} MIMO detectors.
\end{abstract}
\begin{IEEEkeywords}
 MIMO, integer-forcing, lattice reduction, Minkowski reduction, Hermite-Korkine-Zolotareff reduction, complex Lenstra-Lenstra-Lovasz lattice reduction, linear receivers.
\end{IEEEkeywords}

\section{Introduction}
Modern wireless communication systems use multiple antenna transceivers to achieve capacity gains. It is known that such gain comes at the cost of high decoding complexity at the receiver~\cite{zhan12}. On one extreme, high-complexity joint maximum likelihood (ML) decoders can be used at the receiver to reliably decode the information. On the other extreme, there are well-known \emph{linear receivers} such as the ZF receivers, the MMSE receivers~\cite{Kumar09}, and lattice {\blue reduction-aided} MIMO detectors~\cite{Taherzadeh07-1, matz11, Yao02}, which reduce the complexity of the decoding process with respect to ML decoding, trading off some error performance. The ZF, MMSE, and lattice {\blue reduction-aided} MIMO receivers use the knowledge of the channel state information (CSI)~\cite{Taherzadeh07-1, Yao02} at the receiver. For the ZF and the MMSE receivers, the channel coefficient matrix ${\bf H}$ is used \emph{as it is} to recover the information symbols. For the lattice {\blue reduction-aided} MIMO detectors, the matrix ${\bf H}$ is reduced to an equivalent channel matrix through a lattice reduction algorithm, prior to recovering the information symbols. The main purpose of this channel reduction is to obtain an equivalent channel matrix which ``looks more like an orthogonal matrix'', which in turn is more suitable for component-wise symbol decoding. It is well known that the lattice {\blue reduction-aided} MIMO decoding algorithms achieve the full receive diversity~\cite{matz11, Taherzadeh07-2}, while the ZF and MMSE receivers only provide receive diversity of one for the $n\times n$ MIMO channel~\cite{Aria13}. There are also receiver architectures which ``focus" a lattice basis onto another basis suitable for specific detection such as trellis decoding~\cite{Monteiro11}. For MIMO receivers with near optimum performance, we refer the reader to decoding by sampling~\cite{waterman11}. For various transmitter-side techniques that employ linear precoding schemes based on lattice codes, we refer the reader to \cite{Taherzadeh07-1}, \cite{Yang12} and the references within. Throughout this paper, we only consider the receiver-side techniques with no CSIT for MIMO channels.\\
\indent A new receiver architecture called integer forcing (IF) linear receiver has been recently proposed in~\cite{zhan12,or12,zhan10} to attain higher rates in MIMO channels with reduced decoding complexity. In this framework, the source employs a layered transmission scheme, and transmits independent codewords simultaneously across the layers. A distinctive property of this scheme is the use of identical lattice codes as codebooks for each layer. At the receiver side, each layer is allowed to decode an integer linear combination of transmitted codewords. Since any integer linear combination of lattice points is another lattice point, the decoded point will be another lattice point, and this substantially reduces the decoding complexity. This idea for MIMO detection was derived from the compute-and-forward approach for physical layer network coding~\cite{Feng, Liew11, Nazer11, Narayanan10, sakzad12}. In the MIMO IF architecture, a filtering matrix $\textbf{B}$ is used to approximate the channel matrix $\textbf{H}$ to a ``nearest'' integer matrix ${\bf A}$~\cite{zhan12}. In such a case, finding a non-singular integer matrix ${\bf A}$, whose rows play the role of the coefficients of the linear system of equations, is crucial. Hence, a matrix ${\bf B}$ is needed such that $\textbf{A}$ is full rank and ${\bf B}{\bf H} \approx {\bf A}$, with minimum quantization error at high signal-to-noise ratio (${\small \mbox{SNR}}$) values. {\blue In the special case, when ${\bf H}$ has rational entries, the matrix ${\bf B}$ is simply a constant (which is the least common divisor of all the denominators of the entries of ${\bf H}$) times identity matrix.} The problem of finding ${\bf A}$ and ${\bf B}$ for IF receivers is addressed in~\cite{zhan12}. However, its solution is based on an exhaustive search with high computational complexity. In particular, the exhaustive search is computationally complex even for $2\times 2$ real MIMO channel, and becomes impractical to implement for $2\times2$ complex MIMO and higher order MIMO channels.\\
\indent In this paper, we propose a practical method for choosing the integer matrix ${\bf A}$. Our method is based on the HKZ and Minkowski lattice reduction algorithms~\cite{korkine1873, Minkowski1891}, which were recently developed and employed as part of lattice {\blue reduction-aided} MIMO detectors in~\cite{zhang12}. In particular, we use these algorithms to find a matrix ${\bf A}$, which is not only full rank but also unimodular and invertible over the underlying alphabet $\mathcal{R}$. For the $2 \times 2$ and $4 \times 4$ MIMO channels, we compare the performance (in terms of ergodic rate and probability of error) of the proposed practical IF solutions based on HKZ and Minkowski lattice reduction with the known linear receivers. We show that ({\em i}) it provides a lower bound on the ergodic rate of the IF receiver, ({\em ii}) it attains full receive diversity, ({\em iii}) it outperforms lattice reduction-aided detectors in error performance, and ({\em iv}) it trades-off error performance for computational complexity in comparison with the IF receiver based on exhaustive search. Since CLLL has much lower complexity in comparison to HKZ and Minkowski lattice reduction algorithms, we also study the use of CLLL algorithm to find matrix ${\bf A}$. {\blue It has been shown in~\cite{Sakzad13-1} that CLLL-SVD algorithm does not provide full diversity. In this work, we give a new low-complexity method based on CLLL algorithm which achieves full receive diversity only in $2\times 2$ MIMO channels. This algorithm provides the same performance as that of HKZ and Minkowski with much lower complexity only in $2\times 2$ MIMO channels. However, we recall the better performance of IF receivers based on HKZ and Minkowski lattice reduction algorithms for $4 \times 4$ MIMO channels and beyond.} We also provide the connection between IF linear receivers and lattice reduction-aided MIMO detectors, and point out the benefits of the former class of receivers over the latter.\\
\indent The rest of the paper is organized as follows. In Section II, we review the background on lattice reduction algorithms. We present the problem statement along with the signal model in Section III. In Section IV, we study the solution to the IF receiver problem based on three lattice reduction algorithms. The definition of the ergodic rate is introduced and a lower bound on the ergodic rate of IF receiver is also presented. In Section V, we point out the differences between the proposed solution to IF receivers and the lattice {\blue reduction-aided} MIMO detectors. In Section VI, we present simulation results on the ergodic rate and the error performance of IF receiver, and compare these results with the traditional linear receivers as well as the lattice {\blue reduction-aided} MIMO detectors. Finally, we present concluding remarks in Section VII.\\
\indent {\em Notations}. Boldface letters are used for row vectors, and capital boldface letters for matrices. We let $\mathbb{C}$, and $\mathbb{Z}[i]$ denote the field of complex numbers, and the ring of Gaussian integers, respectively, where $i^2 = -1$. We let ${\bf I}_n$ and ${\bf 0}_n$ denote the $n\times n$ identity matrix and zero matrix and the operations $(\cdot)^T$ and $(\cdot)^H$ denote transposition and Hermitian transposition. We let $|\cdot|$ and $\| \cdot \|$ denote the absolute value of a real number and the Euclidean norm of a vector and the operation $\mathbb{E}(\cdot)$ denotes mean of a random variable. We let $\lfloor x \rceil$ and $\lfloor {\bf v} \rceil$ denote the closest integer to $x$ and the component-wise equivalent operation. We denote the $k \times n$ matrix $\textbf{X}=\left[\textbf{x}_{1}^T,\ldots,\textbf{x}_{k}^T\right]^T$, formed from stacking the $n$-dimensional row vectors $ \{ \textbf{x}_{m} ~|~1\leq m\leq k \}$. The symbol $\textbf{X}_{j, m}$ denotes the element in the $j$-th row and $m$-th column of $\textbf{X}$. The notation $\mbox{diag}(x_1,\ldots,x_n)$ refers to a diagonal matrix with entries $x_1,\ldots,x_n$ on its main diagonal and zero elsewhere. If ${\bf X}=\mbox{diag}(x_1,\ldots,x_n)$, then ${\bf X}^r$, for $r\in\mathbb{R}$ denotes $\mbox{diag}(x_1^r,\ldots,x_n^r)$. The Hermitian product of two vectors ${\bf v}$ and ${\bf w}$ is denoted by $\langle{\bf v}, {\bf w}\rangle \triangleq {\bf v}{\bf w}^{H}$. The orthogonal vectors generated by the Gram-Schmidt orthogonalization procedure are denoted by $\{\mbox{GS}({\bf b}_1),\ldots,\mbox{GS}({\bf b}_n)\}$ which spans the same space of $\{{\bf b}_1,\ldots,{\bf b}_n\}$. We define
$$\mu_{m,j}\triangleq\frac{\langle\mbox{GS}({\bf b}_m),\mbox{GS}({\bf b}_j)\rangle}{\|\mbox{GS}({\bf b}_j)\|^2},$$ where $1\leq m,j\leq n$.
\section{Background on Lattices and Lattice Reductions}
A $k$-dimensional {\em lattice} $\Lambda$ with a basis set $\{{\boldsymbol\ell}_1,\ldots,{\boldsymbol\ell}_k\}\subseteq\mathbb{R}^d$ is the set of all points of the form $\{{\bf x}={\bf u}{\bf L}| {\bf u}\in \mathbb{Z}^k\}$ where ${\bf L}$ is the {\em generator matrix} of $\Lambda$, formed by placing ${\boldsymbol\ell}_m$'s as its rows. Throughout the paper, we only consider full rank lattices where $d=k$. The {\em Gram matrix} of $\Lambda$ is ${\bf G}={\bf L}{\bf L}^T$. The {\em $m$--th successive minima} of a lattice, denoted by $\lambda_m$, is the radius of the smallest possible closed ball around origin containing $m$ or more linearly independent lattice points forming a basis. Given a basis set, a lattice reduction technique is a process to obtain a new basis set of the lattice with shorter vectors. Specifically, for the generator matrix ${\bf L}$, the matrix ${\bf L}'={\bf U}{\bf L}$ denotes a reduced generator matrix obtained through a lattice reduction technique, where ${\bf U}$ is a unimodular matrix.
\begin{itemize}
\item{} A lattice generator matrix ${\bf L}'$ is called {\em Minkowski-reduced} if for $1\leq m\leq d$, the vectors ${\boldsymbol\ell}'_m$ are as short as possible~\cite{Minkowski1891}. In particular, ${\bf L}'$ is Minkowski-reduced if for $1\leq m\leq d$, the row vector ${\boldsymbol\ell}_m'$ has minimum possible energy amongst all the other lattice points such that $\{{\boldsymbol\ell}_1',\ldots,{\boldsymbol\ell}_m'\}$ can be extended to another basis of $\Lambda$.
\item{} A generator matrix ${\bf L}'$ for a lattice $\Lambda$ is called {\em HKZ-reduced}~\cite{korkine1873} if it satisfies
\begin{enumerate}
\item{} $|\mu_{m,j}|\leq 1/2$ for all $1\leq j < m\leq d$,
\item{} the vector ${\boldsymbol\ell}'_1$ be the shortest vector of $\Lambda$, and
\item{} the orthogonal projections of the vector ${\boldsymbol\ell}'_1$ onto $\{{\boldsymbol\ell}'_2,\ldots,{\boldsymbol\ell}'_d\}$ is a HKZ-reduced basis.
\end{enumerate}
\item{} A generator matrix ${\bf L}'$ for a lattice $\Lambda$ is called {\em LLL-reduced}~\cite{LLL82} if it satisfies
\begin{enumerate}
  \item $|\mu_{m,j}|\leq 1/2$ for all $1\leq j < m\leq d$, and
  \item $\delta\|\mbox{GS}\left({\boldsymbol\ell}'_{m-1}\right)\|^2\leq \|\mbox{GS}\left({\boldsymbol\ell}'_{m}\right)+\mu_{m,m-1}^2\mbox{GS}\left({\boldsymbol\ell}'_{m-1}\right)\|^2$ for all $1< m\leq d$,
\end{enumerate}
where $\delta \in (1/4, 1]$ is a factor selected to achieve a good quality-complexity tradeoff.
\end{itemize}
For each $1 \leq m \leq d$, it is known that the length of the $m$-th row vector in ${\bf L}'$ is upper bounded by a scaled version of the $m$-th successive minima of $\Lambda$ \cite{zhang12}. For the Minkowski reduction, we have
\begin{equation}~\label{eq:MinkowskiProperties}
\lambda_m^2\leq\|{\boldsymbol\ell}_m'\|^2\leq\max\left\{1,\left(5/4\right)^{d-4}\right\}\lambda_m^2, \mbox{ for } 1\leq m\leq d.
\end{equation}
For the HKZ reduction~\cite{Lagaris90}, we have
\begin{equation}~\label{eq:HKZProperties}
\frac{4\lambda_m^2}{m+3}\leq\|{\boldsymbol\ell}_m'\|^2\leq\frac{(m+3)\lambda_m^2}{4}, \mbox{ for } 1\leq m\leq d.
\end{equation}
For the LLL reduction~\cite{LLL82}, we have
\begin{equation}~\label{eq:LLLProperties}
\beta^{1-m}\lambda_m^2\leq\|{\boldsymbol \ell}_m'\|^2\leq\beta^{d-1}\lambda_m^2, \mbox{ for } 1\leq m\leq d,
\end{equation}
where $\beta=(\delta-1/4)^{-1}$.

Note that the upper bounds given in \eqref{eq:MinkowskiProperties}--\eqref{eq:LLLProperties} are all scalar multiples of the successive minimas. These scalar multiples are exponential in $d$ for the LLL and the Mikowski reduction algorithms, while polynomial in $d$ for the HKZ reduction algorithm.
\section{System Model and Problem Statement}
\label{sec3}
\begin{figure*}[t]%
  \begin{center}%
\includegraphics[width=11cm]{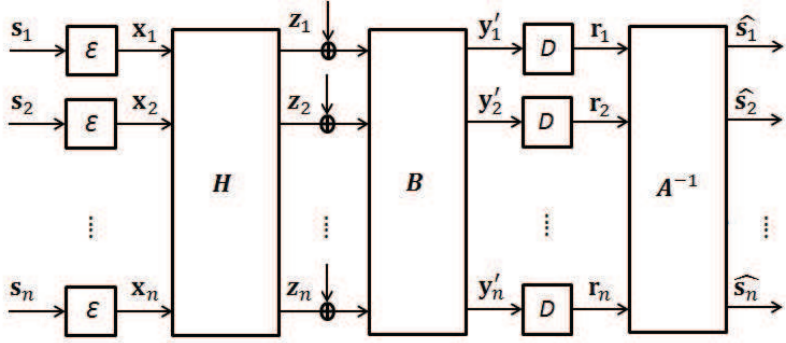}~\caption{\label{fig:bd}
System model block diagram.}
  \end{center}
\end{figure*}
We consider a flat-fading MIMO channel with $n$ transmit antennas and $n$ receive antennas {\blue as Fig.~\ref{fig:bd}}. The channel matrix is denoted by ${\bf H}\in\mathbb{C}^{n\times n}$, where the entries of ${\bf H}$ are i.i.d. as $\mathcal{CN}(0, 1)$. We assume that ${\bf H}$ remains fixed for a given interval (of at least $N$ complex channel uses) and take an independent realization in the next interval. We use a $n$-layer horizontal coding scheme where the information transmitted across different antennas are independent. For $1 \leq m \leq n$, the $m$-th layer is equipped with a lattice encoder $\mathcal{E}:\mathcal{R}^{k}\rightarrow\mathbb{C}^{N}$ which maps a message ${\bf s}_m\in \mathcal{R}^{k}$ over the ring $\mathcal{R}$ into a lattice codeword ${\bf x}_m \in \Lambda \subset \mathbb{C}^{N}$ in the complex space. If ${\bf X}$ denotes the matrix of transmitted vectors, the received signal ${\bf Y}$ is given by
$${\bf Y}_{n\times N} = \sqrt{P}{\bf H}_{n\times n}{\bf X}_{n\times N}+{\bf Z}_{n\times N},$$
where $P=\frac{{\small \mbox{SNR}}}{n}$ and ${\small\mbox{SNR}}$ denotes the average signal-to-noise ratio at each receive antenna. We assume that the entries of ${\bf Z}$ are i.i.d. as $\mathcal{CN}(0, 1)$. We also assume that ${\bf H}$ is known only at the receiver. The goal is to project ${\bf H}$ (by left multiplying it with a receiver filtering matrix ${\bf B}$) onto a non-singular integer matrix ${\bf A}$. In order to uniquely recover the information symbols, the matrix ${\bf A}$ must be invertible over the ring $\mathcal{R}$. Thus, we have
\begin{equation}~\label{eq:LRmodel}
{\bf Y}'\triangleq {\bf B}{\bf Y}=\sqrt{P}{\bf B}{\bf H}{\bf X}+{\bf B}{\bf Z}.
\end{equation}
The above signal model is applicable to all linear receivers including the ZF, MMSE ({\blue with ${\bf A}={\bf I}$ in both cases}), and IF (where ${\bf A}$ is invertible over $\mathcal{R}$).
For the IF receiver~\cite{zhan12} formulation, a suitable signal model is
\begin{equation}~\label{eq:IFmodel}
{\bf Y}'=\sqrt{P}{\bf A}{\bf X}+\sqrt{P}({\bf B}{\bf H}-{\bf A}){\bf X}+{\bf B}{\bf Z},
\end{equation}
where $\sqrt{P}{\bf A}{\bf X}$ is the desired signal component, and the effective noise is $\sqrt{P}({\bf B}{\bf H}-{\bf A}){\bf X}+{\bf B}{\bf Z}$. In particular, the effective noise power along the $m$-th row of ${\bf Y}'$ is defined as
\begin{equation}~\label{quntizederrplusnoise}
g({\bf a}_m,{\bf b}_m)\triangleq P\|{\bf b}_m{\bf H}-{\bf a}_m\|^2 + \|{\bf b}_m\|^2,
\end{equation}
where ${\bf a}_m$ and ${\bf b}_m$ denotes the $m$-th row of ${\bf A}$ and $\bf{ B}$, respectively. Note that in order to increase the effective signal-to-noise ratio for each layer, the term $g({\bf a}_m,{\bf b}_m)$ has to be minimized for each $m$ by appropriately selecting the matrices ${\bf A}$ and ${\bf B}$. We formally put forth the problem statement below:
\vspace{.5cm}
\begin{mdframed}
\begin{Problem}~\label{problem}
Given ${\bf H}$ and $P$, the problem is to find the matrices ${\bf B} \in \mathbb{C}^{n \times n}$ and ${\bf A} \in \mathbb{Z}[i]^{n \times n}$ such that:
\begin{itemize}
\item The $\max_{1\leq m\leq n}g({\bf a}_m,{\bf b}_m)$ is minimized, and
\item The corresponding matrix ${\bf A}$  is invertible over the ring $\mathcal{R}$.
\end{itemize}
\end{Problem}
\end{mdframed}
\vspace{.2cm}
In~\cite{zhan12}, the authors considered the invertibility of the matrix ${\bf A}$ only over finite fields, in which case is equivalent to requiring $\det({\bf A})\neq0$. {\blue A block diagram of the scheme is presented in Fig.~\ref{fig:bd} where $D$ is a lattice decoder which converts ${\bf y}'_m$ to ${\bf r}_{m}$, an estimate for the $m$-th layer of ${\bf A}{\bf X}$.}
\subsection{Known Approaches}
If we choose ${\bf B}={\bf H}^{-1}$ (or pseudo-inverse of ${\bf H}$ in general) and ${\bf A}={\bf I}_n$, then we get the ZF receiver~\cite{Kumar09}. If we choose ${\bf B}={\bf H}^H{\bf S}^{-1}$ (see (4) in \cite{Aria13}), where
\begin{equation}
\label{s_matrix}
{\bf S}=P^{-1}{\bf I}_n+{\bf H}{\bf H}^H,
\end{equation}
and ${\bf A}={\bf I}_n$, then we get the linear MMSE receiver. With this, the well known linear receivers can be viewed under the umbrella of the IF architecture. However, the ZF and MMSE receivers are known not to minimize $g({\bf a}_m,{\bf b}_m)$~\cite{zhan12}. The lattice {\blue reduction-aided} MIMO detector~\cite{Taherzadeh07-1} is another approach which will be discussed in details in Section~\ref{sec5}. {\blue Henceforth, we use ${\bf W}\triangleq{\bf H}^H{\bf S}^{-1}$ to simplify notation.}

\indent In \cite{zhan12}, the authors have proposed a method to solve Problem~\ref{problem}. We now recall the approach presented in \cite{zhan12}. First, conditioned on a fixed ${\bf a}_m = \bf{a}$, the term $g({\bf a},{\bf b}_m)$ is minimized over all possible values of ${\bf b}_m$. As a result, the optimum value of ${\bf b}_m$ can be obtained as
\begin{equation}
\label{b_vector}
{\bf b}_m={\bf a}{\bf H}^H{\bf S}^{-1}.
\end{equation}
Then, after replacing ${\bf b}_m$ from \eqref{b_vector} in $g({\bf a},{\bf b}_m)$, the term $g({\bf a},{\bf a}{\bf H}^H{\bf S}^{-1})$ is minimized over all possible values of ${\bf a}$ to obtain ${\bf a}_{m}$ as ${\bf a}_{m} = \arg \min_{{\bf a}} g({\bf a},{\bf a}{\bf H}^H{\bf S}^{-1})$. {\blue Based on the proof of Theorem 3 in~\cite{zhan12}, the previous expression can be written as
\begin{eqnarray}
{\bf a}_{m} &=& \arg \min_{{\bf a}\in\mathbb{Z}[i]^n} ~{\bf a}\left({\bf I}_n-{\bf H}^H{\bf S}^{-1}{\bf H}\right){\bf a}^H. \label{svd}
\end{eqnarray}
If we replace the singular value decomposition (SVD) of ${\bf H}$ into \eqref{svd}, we get
\begin{eqnarray}
{\bf a}_{m} &=&\arg \min_{{\bf a}\in\mathbb{Z}[i]^n} ~{\bf a}{\bf V}{\bf D}{\bf V}^H{\bf a}^H, \label{opt_problem}
\end{eqnarray}
where ${\bf V}$ is the matrix composed of the eigenvectors of ${\bf H}{\bf H}^{H}$, and ${\bf D}$ is a diagonal matrix with $m$-th entry ${\bf D}_{m,m}=\left(P\rho_m^2+1\right)^{-1}$, where $\rho_m$ is the $m$-th singular value of ${\bf H}$. Since the matrix ${\bf I}_n-{\bf H}^H{\bf S}^{-1}{\bf H}={\bf V}{\bf D}{\bf V}^H$ is symmetric, ${\bf G}={\bf V}{\bf D}{\bf V}^H$ can also be viewed as SVD of ${\bf G}$. This makes ${\bf G}= {\bf L}{\bf L}^H$ a Gram matrix for a lattice with generator matrix ${\bf L}$.} With this, we have to obtain $n$ vectors ${\bf a}_{m}$, $1\leq m \leq n$, which result in the first $n$ smaller values of ${\bf a}{\bf V}{\bf D}{\bf V}^H{\bf a}^H$ along with the non-singular property on ${\bf A}$. In order to get ${\bf a}_m$, $1\leq m\leq n$, the authors of~\cite{zhan12} have suggested an exhaustive search for each component of ${\bf a}_m$
within a sphere of squared radius
\begin{equation}\label{radius}
1+P\rho_{\max}^2,
\end{equation}
where $\rho_{\max}=\max_{m}\rho_m$. Hence, for a fixed $P$, the complexity of this approach is of order $O\left(P^n\right)$. It has also been pointed out in~\cite{zhan12} that this search can be accelerated by {\blue means of a sphere decoder} on the lattice with Gram matrix ${\bf G} = {\bf V}{\bf D}{\bf V}^H$, see~\cite{Wei12}. It is also shown in~\cite{zhan12} that the exhaustive search approach provides a diversity order of $n$ and a multiplexing gain of $n$. At this stage, we note that the exhaustive computation of ${\bf a}_m$ has high complexity, especially for large values of $P$ and $n$, and hence the approach in \cite{zhan12} is not practical even for the $2\times 2$ complex case.

\section{Practical Integer-Forcing MIMO Receivers}
In this section, we propose a practical method to get an invertible integer matrix ${\bf A}$ solving the Problem~\ref{problem}. Once we obtain ${\bf A}$, {\blue we construct} ${\bf B}={\bf A}{\bf \blue W}$, where {\blue ${\bf W}\triangleq{\bf H}^H{\bf S}^{-1}$ for simple notation.} Henceforth, we only address the method for finding ${\bf A}$.
\subsection{Lattice Reduction Algorithms for IF Architecture}
It is pointed out in \cite{zhan12} that the minimization problem in \eqref{opt_problem} is the shortest vector problem for a lattice with Gram matrix ${\bf G}= {\bf V}{\bf D}{\bf V}^H$. Since ${\bf G}$ is a symmetric positive definite matrix, we can write ${\bf G} = {\bf L}{\bf L}^H$ for some ${\bf L} \in \mathbb{C}^{n \times n}$ by using Cholesky decomposition. With this, the rows of ${\bf L}={\bf V}{\bf D}^{\frac{1}{2}}$ generate a lattice, say $\Lambda$. Based on \eqref{opt_problem}, a set of possible choices for $\{{\bf a}_1, \ldots, {\bf a}_n \}$ is the set of complex integer vectors, whose corresponding lattice points in $\Lambda$ have lengths at most equal to the $n$-th successive minima of $\Lambda$.
The two well-known lattice reduction algorithms satisfying the above property up to constants are HKZ and Minkowski lattice reduction algorithms. In particular, we use HKZ and Minkowski lattice reduction algorithms, given in~\cite{zhang12}, to reduce the basis set in ${\bf L}$ and obtain a new generator matrix ${\bf L}'$. Hence, both the HKZ and the Minkowski reduction algorithm can be employed and the rows of ${\bf L}'{\bf L}^{-1}$ can be used to obtain the desired matrix ${\bf A}$ for IF architecture. Since both ${\bf L}'$ and ${\bf L}$ are generator matrices for the same lattice $\Lambda$, the integer matrix ${\bf A}$ is not only invertible over every non-trivial ring but also unimodular.
We summarize this {\blue in {\bf procedure} ALGORITHM1$(H, P)$}, where $\mbox{HKZ}(\cdot)$ and $\mbox{MINKOWSKI}(\cdot)$ refer to HKZ and Minkowski lattice reduction algorithms, respectively whose pseudo-codes are given in~\cite{zhang12}.
\begin{algorithmic}[1]
\Procedure{Algorithm1}{${\bf H},P$}
\State ${\bf S}\gets P^{-1}{\bf I}_n+{\bf H}{\bf H}^H$
\State $({\bf U},\Sigma,{\bf V})\gets\mbox{SVD}({\bf H})$ where $\Sigma=\mbox{diag}(\rho_1,\ldots,\rho_n)$\Comment{The SVD of ${\bf H}$.}
\State ${\bf D}\gets \mbox{diag}\left(\left(P\rho_1^2+1\right)^{-1},\ldots,\left(P\rho_n^2+1\right)^{-1}\right)$
\State ${\bf L}\gets {\bf V}{\bf D}^{\frac{1}{2}}$
\State ${\bf L}'\gets\mbox{HKZ}({\bf L})~(\mbox{or}~\mbox{MINKOWSKI}({\bf L}))$\Comment{Lattice reduction algorithm.}
\State \textbf{return} ${\bf A}={\bf L}'{\bf L}^{-1}$ and ${\bf B}={\bf A}{\bf \blue W}$\Comment{$n$ rows ${\bf a}_m$ of ${\bf A}$ for $1\leq m\leq n$.}
\EndProcedure
\end{algorithmic}

\indent For fixed $P$ and $n$, the expected asymptotic complexity of Minkowski lattice reduction algorithm is upper bounded by
$\left(5/4\right)^{2n^2}$, while the computational complexity of HKZ lattice reduction is of order $(2\pi e)^{n+O(\log 2n)}$ \cite{zhang12}. Note that unlike the IF receiver based on exhaustive search, the complexity of the lattice reduction techniques are independent of $P$. Hence, the above algorithms have lower complexity than the exhaustive search and its accelerated version by sphere decoder \cite{ViB,zhan12}, especially for large values of $P$. Based on \eqref{eq:MinkowskiProperties} and \eqref{eq:HKZProperties}, if $n = 2$ in complex setting (which implies $d = 4$ in \eqref{eq:MinkowskiProperties}), then the Minkowski algorithm provides us a basis set with lengths exactly equal to all successive minimas. For larger values of $n$, the bounds become loose. The following theorem states that the above mentioned loose bound will not reduce the diversity order of the scheme for large values of $n$.
\begin{Theorem}
For a MIMO channel with $n$ transmit and $n$ receive antennas over a Rayleigh fading channel, the integer-forcing linear receiver based on lattice reduction achieves full receive diversity.
\end{Theorem}
\begin{IEEEproof}
The proof is similar to the proof of Theorem $4$ in the extended version of~\cite{zhan12} except that equation (197) in \cite{zhan12} becomes an inequality.
\end{IEEEproof}

\subsection{CLLL Algorithm for IF Receiver}~\label{section:newclll}
One can also consider complex version of the LLL lattice reduction algorithm~\cite{CLLL09} (CLLL) to solve Problem~\ref{problem}. However, since CLLL aims only at obtaining the shortest vector in the corresponding basis, it is not a suitable choice for our problem. In addition, a small improvement to the poor performance of CLLL for IF receiver has been obtained by combining it with two other algorithms~\cite{Sakzad13-1}. This combination of three algorithms is called ``combined CLLL-SVD''. The simulation results show that the combined CLLL-SVD algorithm fails to achieve full receive diversity~\cite{Sakzad13-1}. Now, we proceed to introduce a small modification to Algorithm 2 of ~\cite{Sakzad13-1} in order to get a better quality ${\bf A}$ matrix. Let us define
\begin{eqnarray*}
f({\bf c}, {\bf d}) &\triangleq& P^{-(n+1)}{\bf d}{\bf d}^H+P^{-n}{\bf d}{\bf H}{\bf H}^H{\bf d}^H\nonumber\\
&&-\sqrt{P^{-n}}{\bf c}{\bf H}^H{\bf d}^H-\sqrt{P^{-n}}{\bf d}{\bf H}{\bf c}^H+{\bf c}{\bf c}^H\nonumber,
\end{eqnarray*}
for ${\bf 0}\neq{\bf c}\in\mathbb{Z}[i]^n$ and ${\bf 0}\neq{\bf d}\in\mathbb{C}^n$.
\begin{lemma}
For a fixed ${\bf c}$, the optimum value ${\bf d}_{\mbox{\tiny opt}}$ that minimizes $f({\bf c}, {\bf d})$ is ${\bf d}_{\mbox{\tiny opt}}=\sqrt{P^{n}}{\bf c}{\bf \blue W}$.
\end{lemma}
\begin{IEEEproof}
Setting $0=\frac{\partial f}{\partial {\bf d}}$, we have
\begin{eqnarray*}
\frac{\partial f}{\partial {\bf d}}&=&P^{-(n+1)}{\bf d}^H+P^{-n}{\bf H}{\bf H}^H{\bf d}^H-\sqrt{P^{-n}}{\bf H}{\bf c}^H,\nonumber\\
&=&0,\nonumber
\end{eqnarray*}
which turns out that ${\bf d}_{\mbox{\tiny opt}}=\sqrt{P^{n}}{\bf c}{\bf \blue W}.$
\end{IEEEproof}
Now if we substitute back ${\bf d}_{\mbox{\tiny opt}}$ into $f({\bf c}, {\bf d})$, we get
\begin{eqnarray*}
f({\bf c}, {\bf d}_{\mbox{\tiny opt}})&=&{\bf c}\left(P^{-1}{\bf \blue W}{\bf \blue W}^H+{\bf \blue W}{\bf H}{\bf \blue W}{\bf H}\right.\nonumber\\
                               &&\left.-{\bf \blue W}{\bf H}-{\bf \blue W}{\bf H}\right){\bf c}^H.\nonumber
\end{eqnarray*}
It is easy to see that $f({\bf c}, {\bf d}_{\mbox{\tiny opt}}) = P^{-1}g({\bf c}, {\bf c}{\bf \blue W}) = P^{-1}{\bf c}{\bf VDV}^{H}{\bf c}^{H}$, where the second equality follows from \eqref{svd}. This implies the following proposition.
\begin{Proposition}
A solution ${\bf 0}\neq{\bf c}\in\mathbb{Z}[i]^n$ with corresponding optimum ${\bf d}=\sqrt{P^{n}}{\bf c}{\bf \blue W}$ to minimize the function $f$ is also a solution with corresponding optimum ${\bf d}={\bf c}{\bf \blue W}$ to the minimization of the function $g$.
\end{Proposition}

\indent Based on the above proposition, we take the approach of minimizing $f({\bf c}, \sqrt{P^{n}}{\bf c}{\bf \blue W})$ rather than $g({\bf c}, {\bf c}{\bf \blue W})$ and instead of conditionally minimizing $f({\bf c}, \sqrt{P^{n}}{\bf c}{\bf \blue W})$, we solve the unconditional minimization of $f({\bf c}, {\bf d})$ with an additional constraint of ${\bf d} \in \mathbb{Z}[i]^n$.
We recognize that minimizing $f({\bf c}, {\bf d})$ with ${\bf c}, {\bf d}\in\mathbb{Z}[i]^n$ is nothing but finding the shortest vector of a $2n-$dimensional complex lattice $\Lambda'$ generated by
\begin{equation}~\label{G}
{\bf L}=\left[\begin{array}{c|c}
\sqrt{P^{-(n+1)}}{\bf I}_n &-\sqrt{P^{-n}}{\bf H}\\
\hline
{\bf 0}& {\bf I}_n
\end{array}\right]  \in \mathbb{C}^{2n \times 2n}.
\end{equation}
The above observation stems from the fact that $f({\bf c}, {\bf d}) = {\bf v}{\bf L}{\bf L}^H{\bf v}^H=\|{\bf v}{\bf L}\|^2$, where ${\bf v} \in \mathbb{Z}[i]^{2n}$ and ${\bf v} = \left[{\bf d}|{\bf c}\right]$ formed by adjoining the vector ${\bf d}$ after ${\bf c}$.

Our CLLL-based solution for Problem \ref{problem} is as follows: for the matrix ${\bf L}$ in \eqref{G}, let ${\bf L}'$ denote the $2n$--dimensional CLLL-reduced generator matrix of $\Lambda'$. Using the short vectors in ${\bf L}'$, we obtain the complex integer matrix ${\bf V}$ with $2n$ row vectors ${\bf v}_{m} = \left[{\bf d}_{m}|{\bf c}_{m}\right]$ such that ${\bf V} = {\bf L}'{\bf L}^{-1}$. It gives us complex integer vectors ${\bf d}_{m}, {\bf c}_{m}$ resulting in smaller values for $f({\bf c}, {\bf d})$ and consequently smaller values for $g({\bf c}, {\bf d})$. Hence, we can use the vectors ${\bf d}_{m}, {\bf c}_{m}$ for the IF architecture as ${\bf b}_{m} = {\bf d}_{m}$, and ${\bf a}_{m} = {\bf c}_{m}$. Note that we ignore ${\bf b}_{m}$ as it is a complex integer vector and we only use ${\bf a}_{m}$, and subsequently, obtain ${\bf b}_{m}$ using \eqref{b_vector}. {\blue A summary of this can be found in {\bf procedure} ALGORITHM2$(H, P)$}, where $\mbox{CLLL}(\cdot)$ refers to the CLLL algorithm whose pseudo-code is given in~\cite{CLLL09}.
\begin{algorithmic}[1]
\Procedure{Algorithm2}{${\bf H},P$}
\State ${\bf S}\gets P^{-1}{\bf I}_n+{\bf H}{\bf H}^H$
\State ${\bf L}\gets \left[\begin{array}{c|c}
\sqrt{P^{-(n+1)}}{\bf I}_n &-\sqrt{P^{-n}}{\bf H}\\
\hline
{\bf 0}& {\bf I}_n
\end{array}\right]$
\State ${\bf L}'\gets \mbox{CLLL}({\bf L})$
\State ${\bf V} \gets {\bf L}'{\bf L}^{-1}$
\State ${\bf v}_{m}\gets\left[{\bf d}_{m} |{\bf c}_{m}\right]$\Comment{$1\leq m\leq 2n$}
\State ${\bf a}_{m}\gets{\bf c}_{m}$ and ${\bf b}_m\gets{\bf a}_m{\bf \blue W}$\Comment{$1\leq m\leq 2n$}
\State \textbf{return} ${\bf A}$ and ${\bf B}={\bf A}{\bf \blue W}$\Comment{The best $n$ vectors satisfying conditions of Problem~\ref{problem}}
\EndProcedure
\end{algorithmic}

\subsection{Ergodic Rate of IF Linear Receivers}
For a given ${\bf H}$ matrix, the achievable rate $R$ under the IF architecture is given by \cite{zhan12},
\begin{equation}~\label{eq:rate1}
R<\min_mnR({\bf H},{\bf a}_m,{\bf b}_m),
\end{equation}
where
\begin{equation}~\label{eq:rate2}
R({\bf H},{\bf a}_m,{\bf b}_m)=\log^+\left(\frac{P}{g({\bf a}_m,{\bf b}_m)}\right),
\end{equation}
is the achievable rate for the $m$--th layer of lattice decoding, and $\log^+(x)=\max\{\log(x),0\}$. Hence, the overall rate is dominated by the layer which has the largest value of $g({\bf a}_m,{\bf b}_m)$. Using \eqref{eq:rate2}, we now define the ergodic rate of the IF architecture for a $n \times n$ MIMO channel as below.
\begin{Definition}~\label{def:er}
The ergodic rate $R_e$ of an IF receiver for a MIMO channel is defined as
$$R_e \triangleq \mathbb{E}\left\lbrace \min_mnR({\bf H},{\bf a}_m,{\bf b}_m)\right\rbrace,$$
where the mathematical expectation is taken over the channel coefficient matrix ${\bf H}$.
\end{Definition}

\indent Since ${\bf a}_m$ and ${\bf b}_m$ are functions of ${\bf H}$, we can alternatively
denote $R({\bf H},{\bf a}_m,{\bf b}_m)$ as
$R({\bf H},{\bf a}_m({\bf H}),{\bf b}_m({\bf H}))$.
As a result, the definition of the ergodic rate does not depend on
a specific matrix pair ${\bf A}$ and ${\bf B}$.
For a given $\textbf{H}$ and $P$, let $\{ \overline{{\bf a}}_m, \overline{{\bf b}}_m ~|~ 1 \leq m \leq n \}$ denote the IF solution based on the lattice reduction algorithm. Further, let $R_e'$ denote the corresponding ergodic rate which is obtained as per Definition \ref{def:er}. It is straightforward to observe that the ergodic rate of the IF architecture is lower bounded by $R_e'$, since $\{ {\bf a}_m, {\bf b}_m ~|~ 1 \leq m \leq n \}$ obtained using the exhaustive search results in smaller values of $g(\cdot, \cdot)$ when compared to the lattice reduction technique.

\section{Comparison of IF receivers with Lattice Reduction Aided MIMO Detectors}
\label{sec5}
In this section, we point out the differences between the conventional lattice {\blue reduction-aided} MIMO detectors, and the IF receivers based on lattice reduction techniques. {\color{black} In \cite{Taherzadeh07-2}, the authors have analyzed two types of lattice reduction-aided decoding based on reducing either dual or primal lattice, and have shown that the former method is more appropriate to reduce the effective noise, see also~\cite{Ling11, Monteiro11-2}. Hence, we only compare IF receivers with lattice reduction-aided detectors, which use dual lattice base reduction. In this technique, the goal is to reduce the dual lattice generator matrix ${\bf H}^{-H}$ to an equivalent matrix ${\bf H}'$ using the lattice reduction algorithms~\cite{zhan12, matz11, CLLL09}. If the columns of ${\bf H}'$ denote the reduced basis set corresponding to the columns of ${\bf H}^{-H}$, then we get ${\bf H}' = {\bf H}^{-H}{\bf U},$ where ${\bf U}$ is a unimodular matrix. If we use
\begin{equation}
\label{lr_zf}
{\bf B} = \left({\bf H}'\right)^H = {\bf U}^H{\bf H}^{-1},
\end{equation}
and ${\bf A} = {\bf U}^H$, then \eqref{eq:LRmodel} can be written as ${\bf Y}'=\sqrt{P}{\bf U}^H{\bf X}+{\bf B}{\bf Z}$.} Since {\blue ${\bf U}^H$} is a unimodular matrix, it is invertible over $\mathcal{R}$ and the information symbols can be recovered by solving a system of linear equations based on {\blue ${\bf U}^H$}. Henceforth, when we use lattice reduction on {\blue ${\bf H}^{-H}$} and use {\blue ${\bf B}= {\bf U}^H{\bf H}^{-1} $}, we refer to such a method as ``lattice reduction zero-forcing'' (LR-ZF) detector. Apart from the above choice of ${\bf B}$, one can also use
{\blue
\begin{equation}
\label{lr_mmse}
{\bf B} = {\bf U}^H{\bf \blue W},
\end{equation}
}
as in \eqref{b_vector} to obtain a better projection matrix. For such a choice, we refer to the method as  ``lattice reduction MMSE'' (LR-MMSE) detector. Note that for large values of $P$, the matrix ${\bf \blue W}$ is the pseudo-inverse of ${\bf H}$, and hence, the LR-ZF detector and the LR-MMSE detector are the same~\cite{jiang11}. It follows that LR-MMSE is a natural generalization of LR-ZF. We are interested in LR-MMSE since MMSE is known to perform better than ZF at low and moderate $P$ values.\\
\indent We now compare the lattice {\blue reduction-aided} detectors with the IF linear receiver based on lattice reduction techniques. To facilitate the comparison, we study the role of LR-ZF and LR-MMSE detectors in solving the Problem~\ref{problem}. Along that direction, applying lattice reduction on {\blue ${\bf H}^{-H}$} can be viewed as the result of substituting ${\bf b}_m={\bf a}{\bf H}^{-1}$ in the problem of minimizing $g({\bf a}_m,{\bf b}_m)$ conditioned on a fixed ${\bf a}_m = \bf{a}$. However, we already know that the solution to the above conditional minimization problem is given by \eqref{b_vector}, which is not ${\bf b}_m={\bf a}{\bf H}^{-1}$. Hence, the choice of ${\bf B}$ in \eqref{lr_zf} and \eqref{lr_mmse} does not minimize the effective noise. This explains the weakness of the LR-ZF and LR-MMSE receivers in solving the Problem~\ref{problem}, and in-turn explains the benefit of the proposed IF linear receivers. This key difference between the IF receivers and lattice {\blue reduction-aided} detectors can be pointed to Step 1 in Table \ref{IF_LR_comparison}. This difference will result in performance degradation of lattice {\blue reduction-aided} decoders in comparison with IF receivers at low and moderate values of $P$.
However, for large values of $P$, the error performance of LR-ZF, LR-MMSE and the IF receiver based on lattice reduction will coincide since the MMSE solution is known to coincide with the ZF solution at high SNR values. The above advantages are applicable for IF receivers based on lattice reduction techniques. Apart from the above discussed advantages, in general, the IF receiver brings in the following advantages:
({\em i}) in the IF receiver, if the computations are done over a finite field, the integer matrix ${\bf A}$ must be non-singular, however, in the lattice {\blue reduction-aided} detectors, the integer matrix ${\bf U}$ is unimodular, which is a stronger condition than the non-singularity property. This relaxation in the constraint will help the IF receivers in selecting a better integer matrix ${\bf A}$, ({\em ii}) the other difference, pointed out in \cite{zhan12}, comes from the level of operation of the decoder. The well-known lattice {\blue reduction-aided} detectors work at symbol-level by detecting the symbols of {\blue ${\bf U}^H{\bf X}$} from the received vectors. However, the IF receiver primarily works at codeword level, although one can then look at the symbol level as well.
\begin{small}
\begin{table*}
\caption{The role of IF, LR-ZF, and LR-MMSE detectors in solving the Problem~\ref{problem}. In this table, LR denotes lattice reduction. We recall that ${\bf W}={\bf H}^H{\bf S}^{-1}$.}
\begin{center}
\begin{tabular}{||c|c|c|c||}
\hline\hline
& IF receiver & LR-ZF receiver & LR-MMSE receiver \\
\hline\hline
Step 1 & Substitute $\small{{\bf b}_m={\bf a}{\bf \blue W}}$  &  Substitute ${\bf b}_m={\bf a}{\bf H}^{-1}$ & Substitute ${\bf b}_m={\bf a}{\bf H}^{-1}$ \\
& in $g({\bf a}_m,{\bf b}_m)$ & in $g({\bf a}_m,{\bf b}_m)$ & in $g({\bf a}_m,{\bf b}_m)$\\
\hline
Step 2 & $\arg \min_{{\bf a}\in\mathbb{Z}[i]^n} ~{\bf a}{\bf V}{\bf D}{\bf V}^H{\bf a}^H$ & $\arg \min_{{\bf a}\in\mathbb{Z}[i]^n} ~{\bf a}{\bf H^{-1}}({\bf H^{-1}})^H{\bf a}^H$ & $\arg \min_{{\bf a}\in\mathbb{Z}[i]^n} ~{\bf a}{\bf H^{-1}}({\bf H^{-1}})^H{\bf a}^H$\\
\hline
Step 3 & employ LR on & employ LR on & employ LR on\\
& ${\bf V}{\bf D}^{\frac{1}{2}}$ & {\blue ${\bf H}^{-H}$} & {\blue ${\bf H}^{-H}$}\\
\hline
Step 4 & use the output of LR & use the output of LR & use the output of LR\\
& as the rows of as ${\bf A}$ & {\blue as the rows of ${\bf U}^H$} & {\blue as the rows of ${\bf U}^H$}\\
\hline
Step 5 & use ${\bf B}={\bf A}{\bf \blue W}$ & {\blue use ${\bf B}={\bf U}^H{\bf H}^{-1}$} & {\blue use ${\bf B}={\bf U}^H{\bf \blue W}$} \\
\hline\hline
\end{tabular}
\end{center}
\label{IF_LR_comparison}
\end{table*}
\end{small}

\section{Simulation Results}~\label{sec_4}
In this section, we present simulation results on the ergodic rate and the probability of error of various linear receivers. For the ergodic rate, we compare the IF receivers based on the exhaustive search and the lattice reduction algorithm on $2 \times 2$ MIMO channels. For the probability of error, we compare the following receiver architectures on $2 \times 2$ and $4\times 4$ MIMO channels: ({\em i}) IF linear receiver with exhaustive search, ({\em ii}) IF linear receiver with lattice reduction solutions, ({\em iii}) lattice {\blue reduction-aided} detectors, and ({\em iv}) the joint maximum likelihood (ML) decoder. For the IF receiver with exhaustive search, the results are presented with the constraint of fixed radius for the exhaustive search. We have not used the radius constraint given in \eqref{radius} as the corresponding search space increases with $P$. Instead, we have used a fixed radius of $8$ for all values of $P$. This radius was optimized based on the experimental results by studying the trade-off between the complexity and the error-performance for various radius values.
For the $2 \times 2$ complex MIMO channel, there are only $4$ shortest vectors to be found in the real dimension. Further, the entries of the channel are Gaussian distributed with higher concentration towards the zero value. This intuition prompted us to experiment with shorter values on the radius. With this new radius, we have noticeably reduced the complexity of brute force search.
\subsection{Ergodic Rate Results}
In order to obtain the ergodic rate, we have used \eqref{eq:rate1}, \eqref{eq:rate2} and Definition~\ref{def:er}. To this end, for a given ${\bf H}$ and $P$, the corresponding matrices ${\bf A}$ and ${\bf B}$ are obtained using various introduced methods. Next, we have calculated \eqref{eq:rate2} for each row of ${\bf A}$ and ${\bf B}$. Finally, \eqref{eq:rate1} is computed and an average of \eqref{eq:rate1} over different realizations of ${\bf H}$ gives us the ergodic rate corresponding to the employed algorithm at $P$.
\begin{figure}
  \begin{center}%
\includegraphics[width=9cm]{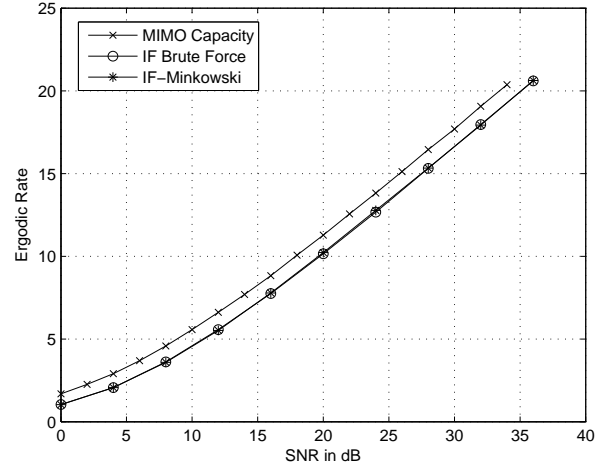}~\caption{\label{fig:ErgodicRateCapacity22}
Ergodic rate of various linear receivers for $2 \times 2$ MIMO channel.}
  \end{center}
\end{figure}
\begin{figure}[t]%
  \begin{center}%
\includegraphics[width=9cm]{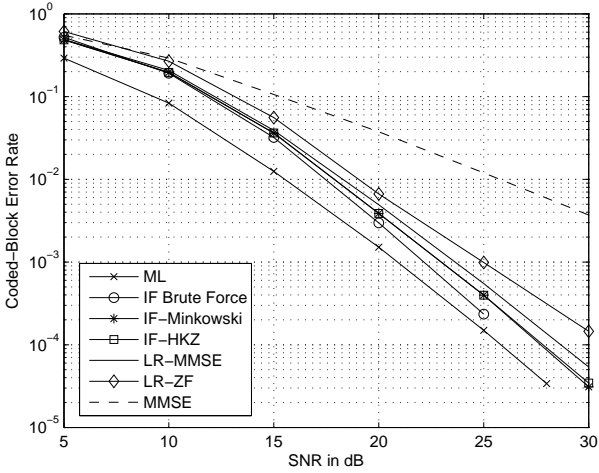}~\caption{\label{fig:FER22}Coded-block error rate for various IF linear receivers versus lattice {\blue reduction-aided} MIMO detectors with $4$-QAM constellation over a $2\times2$ MIMO channel.}
  \end{center}
\end{figure}

In Fig. \ref{fig:ErgodicRateCapacity22}, we compare the ergodic rates of the IF receivers with the ergodic MIMO capacity~\cite{Telatar99} for $2 \times 2$ MIMO channel. Note that the IF receiver based on Minkowski lattice reduction provides ergodic rate approximately same as that of the IF receiver based on exhaustive search. For the IF receiver based on exhaustive search, we have searched for a non-singular ${\bf A}$ over $\mathbb{Z}[i]$. For higher order MIMO channels, we have not studied the tightness of the lower bound since the exhaustive search is too complex to implement.
\subsection{Error Probability Results}
For the error probability results, we employ the points from a $2$-dimensional integer lattice as codewords. With reference to the signal model in Section II, we use $N = 1$ and $\Lambda = \mathbb{Z}[i]$. In general, large block-length integer lattice codes (with $N > 1$) can be used to further improve the error performance. However, in this work, we are only interested in the diversity achieved by the proposed reduction algorithms, and hence, we consider $N = 1$. We now present the error performance of various receiver architectures with $4$-QAM constellation. We use the finite constellation $\mathcal{S} = \{ 0, 1, i, 1+i \}$ carved out of the infinite lattice $\mathbb{Z}[i]$, where $\mathcal{S}$ is the set of coset representatives of $\mathbb{Z}[i]/2\mathbb{Z}[i]$. With reference to the signal model in Section II, $\mathcal{R}$ corresponds to the finite ring $\mathbb{Z}_{2} = \{0, 1\}$. In this method, an appropriate translated version of the symbols of $\mathcal{S}$ is transmitted to reduce the average transmit power and removed at the receiver. After suitable modification on the received vector we get, ${\bf y} = \sqrt{P}{\bf H}{\bf s} + {\bf z}$, where ${\bf s}\in \mathcal{S}^{n \times 1}$. Using the standard one-to-one relation between $\mathbb{C}$ and $\mathbb{R}^{2}$, the received vector ${\bf y}$ is unfolded to a real vector \cite{zhang12} to obtain $\bar{{\bf y}} = \sqrt{P}\bar{{\bf H}}\bar{{\bf s}} + \bar{{\bf z}}$, where $\bar{{\bf s}} \in \{ 0, 1\}^{2n \times 1}$. For this setting, we use modulo lattice decoding at the receiver as follows:
\begin{enumerate}
\item \textbf{Infinite lattice decoding}: Each component of ${\bf B}\bar{{\bf y}}$ is decoded to the nearest point in $\mathbb{Z}$ to get $\hat{{\bf y}}$. In particular, we use $\hat{{\bf y}} = \lfloor{\bf B}\bar{{\bf y}}\rceil$.
\item \textbf{Projecting onto lattice codewords}: Then, ``mod $2$" operation is performed independently on the components of $\hat{{\bf y}}$. With this, we get ${\bf r} \equiv\hat{{\bf y}} \pmod{2}$.
\item \textbf{Decoupling the lattice codewords}: Further, we solve the system of linear equations ${\bf r} \equiv {\bf A}\bar{{\bf s}} \pmod{2}$ over the ring $\{0, 1\}$ to obtain the decoded vector $\hat{{\bf s}}={\bf A}^{-1}{\bf r} \pmod{2}$.
\end{enumerate}
For higher order constellations such as $M$-QAM, where $M$ is a power of $2$, the decoding process is similar to the above steps except that the second and the third steps work on the finite ring $\mathbb{Z}_{\sqrt{M}} = \left\{0, 1, \ldots, \sqrt{M} \right\}$. In general, for decoding integer lattice codes of large block-length (with $N > 1$), the above listed decoding procedure should be suitably modified based on the structure of the underlying lattice code. Obviously, the complexity of decoding arbitrary length lattices is larger than the rounding operation used for the $2$-dimensional lattices, and in particular, the complexity depends on the existence of a low complexity decoder for the underlying lattice code. In the second and third steps, the complexity grows with the block length $N$. For example, if the lattice code $\mathcal{C} \subset \mathbb{Z}[i]^{16}$ from the Barnes-Wall lattice $BW_{16} = \mathcal{C} + 4\mathbb{Z}[i]^{16}$ \cite{HVB} is employed (which corresponds to $N = 16$), then step 1 is achieved by the decoder in \cite{HVB}, while, step 2 and step 3 are solved over the finite ring $\mathbb{Z}_{4} = \{0, 1, 2, 3 \}$.
{\blue
\begin{rem}
The last step of our decoding process involves inversion (over real numbers) and mod 2 operation. These two operations together can be considered as the inversion operation over the ring $\{0, 1\}$. In our setting, the matrix ${\bf A}$ is unimodular and hence invertible. Since ${\bf A}$ is always a square invertible matrix, the inversion (using the Guassian elimination technique) gives a unique $\bar{{\bf s}}$.
\end{rem}}
\begin{figure}[t]%
  \begin{center}
\includegraphics[width=9cm]{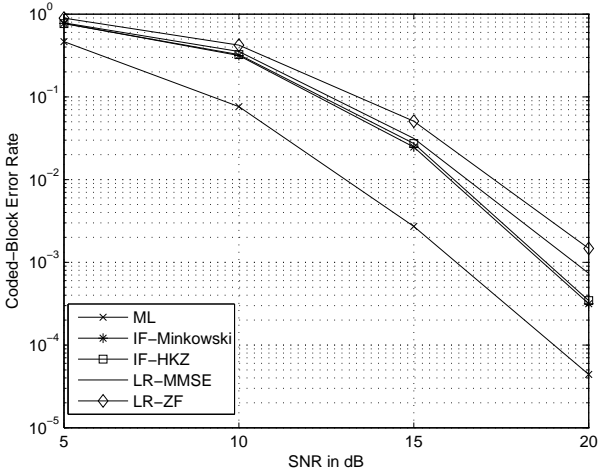}~\caption{\label{fig:FER44}Coded-block error rate for various IF linear receivers versus lattice {\blue reduction-aided} MIMO detectors with $4$-QAM constellation over a $4\times4$ MIMO channel. The IF Brute Force is computationally complex to simulate in this case.}
  \end{center}
\end{figure}
\subsubsection{Coded-Block Error Rate}
We are interested in studying the capability of the proposed solutions in reducing the effective noise in \eqref{eq:IFmodel}. Along that direction, we define a coded-block error if $\hat{{\bf y}}\neq {\bf A}\bar{{\bf s}}$. This refers to the event of incorrectly decoding a block of codewords transmitted across all the antennas. In order to obtain the coded-block error results for the IF receiver, we implement the first step of the above decoding procedure. In Fig. \ref{fig:FER22}--\ref{fig:FER44}, we present the coded-block error rate results for all the receiver architectures. For the IF receiver based on exhaustive search, we search for non-singular ${\bf A}$ over $\mathbb{Z}[i]$. We refer to such a method as ``IF Brute Force''. 
From the figures, note that IF receiver with lattice reduction solutions outperform the MMSE, LR-ZF, and LR-MMSE receivers. This difference in the performance between the proposed IF receiver and the lattice {\blue reduction-aided} MIMO detectors confirm the outcome of our comparative study in Section \ref{sec5}. For the LR-ZF, and LR-MMSE receivers, we use the Minkowski reduction algorithm, and the decoding operation is same as the IF receiver except that the matrix ${\bf B}$ is obtained as in \eqref{lr_zf} and \eqref{lr_mmse}, respectively. Evident from the figures, the proposed lattice reduction solution achieves full receive diversity but trades-off error performance for complexity in comparison with IF brute force search. In the IF receiver based on brute force search, we search for non-singular ${\bf A}$ over $\mathbb{Z}[i]$ (but not necessarily invertible over $\mathbb{Z}[i]$, i.e. ${\bf A}^{-1}\notin\mathbb{Z}[i]^{n\times n}$), however, in the IF receiver based on lattice reduction, the integer matrix ${\bf A}$ is unimodular, which is a stronger condition than the non-singularity property. This relaxation in the constraint can be attributed to the difference in the performance between IF receivers based on exhaustive search and lattice reduction solution. {\blue This has also been pointed out in~\cite{zhan12}.} From the figure, it is clear that the ML decoder outperforms the class of linear receivers. For the proposed IF receiver, a lattice reduction algorithm is performed only at the beginning of each quasi-static interval and subsequently, a system of linear equations has to be solved for each codeword use. However, for the ML decoder, a sphere decoder algorithm is performed for every codeword use. Hence, the complexity of the proposed IF receiver is lower than the ML decoder for slow varying channels.
\begin{figure}[t]%
  \begin{center}
\includegraphics[width=9cm]{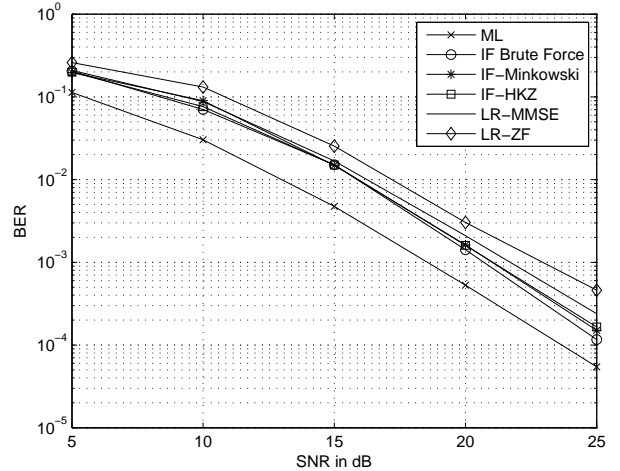}~\caption{\label{fig:BER22}BER for various IF linear receivers versus lattice {\blue reduction-aided} MIMO detectors with $4$-QAM constellation over a $2\times2$ MIMO channel.}
  \end{center}
\end{figure}
\subsubsection{Bit Error Rate}
\begin{figure}[t]%
  \begin{center}
\includegraphics[width=9cm]{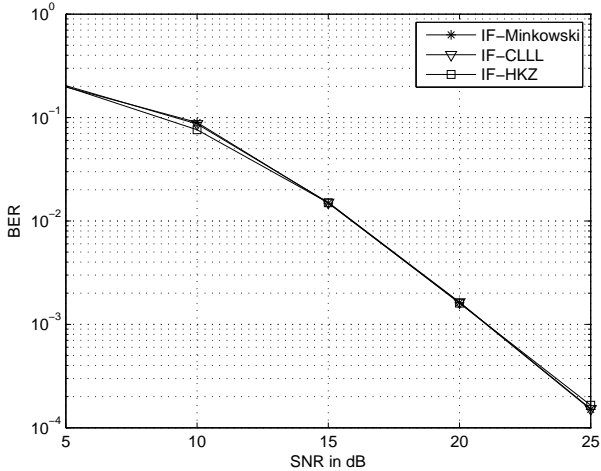}~\caption{\label{BERnew22}BER for various IF linear receivers based on brute force search and various lattice reduction algorithms with $4$-QAM constellation over a $2\times2$ MIMO channel.}
  \end{center}
\end{figure}

\begin{figure}[t]%
  \begin{center}
\includegraphics[width=9cm]{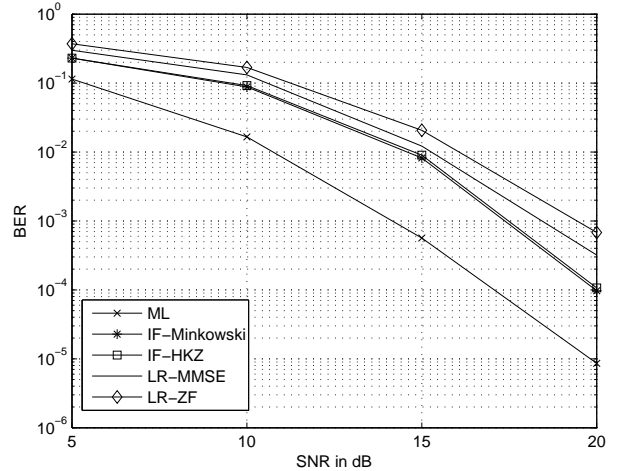}~\caption{\label{fig:BER44}BER for various IF linear receivers versus lattice {\blue reduction-aided} MIMO detectors with $4$-QAM constellation over a $4\times4$ MIMO channel. The IF Brute Force is computationally complex to simulate in this case.}
  \end{center}
\end{figure}

\begin{figure}[t]%
  \begin{center}
\includegraphics[width=9cm]{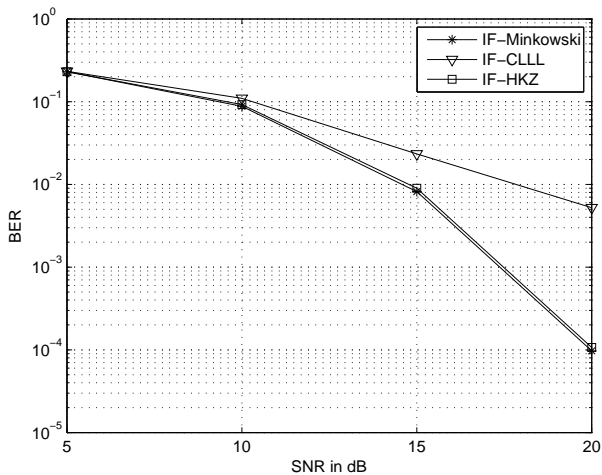}~\caption{\label{fig:BERnew44}BER for various IF linear receivers based on brute force search and various lattice reduction algorithms with $4$-QAM constellation over a $4\times4$ MIMO channel. The IF Brute Force is computationally complex to simulate in this case.}
  \end{center}
\end{figure}

To obtain the BER results of the IF receiver, we have implemented all the steps explained in the decoding procedure. For the IF brute-force search, we have searched for  ${\bf A}$ which is invertible over $\mathcal{S}$. This additional constraint is necessary to solve the linear system of equations with a unique solution.
In Fig. \ref{fig:BER22}--\ref{fig:BERnew44}, we present the BER results for all the receiver architectures. We call the approach based on CLLL algorithm introduced in \ref{section:newclll} as ``IF CLLL''. The figures show that in the case of $2\times 2$ MIMO channel the IF receiver with lattice reduction solution marginally trades-off error performance for complexity in comparison with brute force search. In particular, our approach provides diversity results as that of the exhaustive search approach with much lower complexity in comparison with fixed radius exhaustive search. The error performance of lattice {\blue reduction-aided} MIMO detectors are also presented as other methods of low-complexity detectors which achieve full diversity. Note that for the $4\times4$ MIMO channel the IF-CLLL solution is the only one which fails to provide diversity results.
\subsection{Complexity Comparison}~\label{sec:time}
\begin{figure}[t]%
  \begin{center}
\includegraphics[width=9cm]{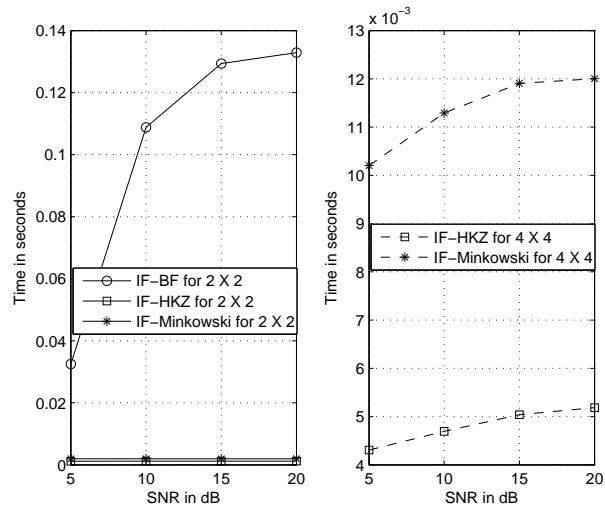}~\caption{\label{fig:time} Average time taken by the HKZ and Minkowski lattice reduction algorithms against brute force search as in~\cite{zhan12} to produce the output for various SNR values. Note that the IF Brute Force for the $4\times4$ configuration is too complex to implement.}
  \end{center}
\end{figure}

To get an idea on the relative complexities among the proposed methods (as they are used for simulations), in Fig.~\ref{fig:time} we provide the average computation time of the three algorithms to produce the output for various $P$ values. These three algorithms are HKZ, Minkowski and the brute force as in~\cite{zhan12}. {\blue We again remind the reader that the base line option in terms of performance is exhaustive search proposed in~\cite{zhan12}.} It is clear from the figure that our solution based on HKZ and Minkowski lattice reductions has significantly lower complexity than the brute force search. {\blue Note also that the average computation time for the brute force search increases with $P$ up to ${\small \mbox{SNR}}=20$dB, since we have used $\min\left\{8,\sqrt{1+P\rho_{\max}^2}\right\}$ as the radius.} 

\indent The complexity and the diversity results of various IF receivers are summarized in Table \ref{complexity_table}. The complexity expressions suggest that Combined CLLL-SVD and CLLL reduction presented in Algorithm 2 has lower complexity in comparison with the HKZ and Minkowski algorithms, especially for higher order MIMO channels.

\begin{table*}
\caption{\label{table2} Summary of results.}
\begin{center}

\begin{tabular}{||c|c|c||} \hline\hline
\multirow{1}{*}Approach & Complexity & Receive diversity\\
\hline\hline
IF based on  Brute Force as in~\cite{zhan12} & $P^n$ & Full\\
IF based on HKZ reduction & $(2\pi e)^{n+O(\log 2n)}$ & Full\\
IF based on Minkowski reduction & $(5/4)^{2n^2}$ & Full\\
IF based on Combined CLLL-SVD as in~\cite{Sakzad13-1} & $(n^4\log{n})/2$ & Not Full\\
IF based on CLLL reduction in Algorithm 2& $((2n)^4\log{2n})/2$ & Not Full (except for $2\times2$)\\
\hline\hline
\end{tabular}
\end{center}
\label{complexity_table}
\end{table*}

\section{Conclusions and Future Directions}
The problem with designing MIMO IF receiver architecture has two folds: ({\em i}) to find the integer matrix $\textbf{A}$ based on the channel matrix $\textbf{H}$ such that the effective noise (after the post-processing operations) is minimized, ({\em ii}) to design efficient lattice codes which when used in the IF architecture improves the error performance of IF receiver. Though the problem statements in ({\em i}) and ({\em ii}) appear independent, the solution to ({\em i}) is key to the effectiveness of the solution in ({\em ii}). Therefore, in this paper, we have proposed a systematic method based on HKZ and Minkowski lattice reduction algorithms to obtain integer coefficients for the MIMO IF architecture. We have also discussed the possible use of CLLL algorithm. 
We have presented the simulation results on the ergodic rate, error performance and the average computation complexity to reveal the effectiveness of lattice reduction solution in comparison with other linear receivers. We have also shown the connections between our solution and the conventional lattice-aided MIMO detectors. In summary, the proposed approach provides full receive diversity at a much lower complexity in comparison with the optimum solution based on exhaustive search. {\blue The IF receiver architecture still has the following limitation. That is, the complexity of finding the matrix ${\bf A}$ using HKZ and Minkowski lattice reduction algorithms is still high when the number of antennas increase.} To verify the receive diversity of the proposed solution, we have employed the $2$-dimensional integer lattice constellations. Once the diversity property with $2$-dimensional constellations (uncoded system) is shown, it is straightforward to see that lattice codes will continue to provide the receive diversity, with additional coding gain due to the distance properties of the lattice code.
In general, a complete solution to the IF architecture demands the design of good lattice codes (of large block lengths) which are not only optimized in terms of error performance but also compatible to work with decoders with lower decoding complexity. This problem of lattice code construction is out of the scope of this paper, and is certainly an important problem for future work. Some possible strong lattices can be found in~\cite{HVB,LDPCLattice,TL, LDA}, and~\cite{LDLC}. Studying coded IF schemes with outer codes such as turbo and LDPC codes is also an interesting forthcoming direction, which could reveal more details on the possibility of near ML performance of the IF receivers. {\blue Comparing IF architecture with other MIMO linear detectors such as lattice reduction algorithm followed by successive interference cancelation (SIC) is also of interest.} Another possible future work is to study the tightness of \eqref{radius}, and possibly propose tighter bounds on this radius.


\end{document}